\documentstyle[12pt]{article}

\def\be{\begin{equation}}
\def\ee{\end{equation}}
\def\bea{\begin{eqnarray}}
\def\eea{\end{eqnarray}}
\def\line{\hbox to \hsize}    
\def\frac #1#2{{#1\over #2}}
\def\Tr{{\rm  Tr\,}}
\def\tr{{\rm  tr\,}}

\def\psid{\psi^{\dagger}}

\def\Z{{\cal Z}}

\def\Det{{\rm Det\,}}

\def \n{{\bf n}}

\def\eval #1#2#3{{\langle#1\vert#2\vert#3\rangle}} 
\def\f1d{{\cal F}^{(1d)}}
\def\1{\mbox{\bf 1}}
\def\l{\overline{\Lambda }}
\def\L{\overline{\underline{\Lambda }}}

\begin{document}

\title{THE GRADIENT EXPANSION FOR THE FREE-ENERGY \\ OF A CLEAN 
SUPERCONDUCTOR}

\author{{\v S}IMON KOS, MICHAEL STONE
\\University of Illinois, Department of Physics\\ 1110 W. Green St.\\
Urbana, IL 61801 USA\\E-mail: m-stone5@uiuc.edu}   

\maketitle

\begin{abstract}We describe a novel method for obtaining the gradient
expansion for the free energy of a clean BCS superconductor. We
present explicit results up to fourth order in the gradients of the
order parameter.

\end{abstract}

\newpage

\section{Introduction}

The Landau-Ginzburg equations \cite{werthamerR} encapsulate much of
the physics of conventional superconductors.  They were originally
proposed on  phenomenological grounds, but after the  advent of
the BCS theory  Gor'kov \cite{gorkov1} provided  a  formal derivation
of the Landau-Ginzburg free energy, valid in the vicinity of $T_c$.
Gor'kov's derivation was soon extended to a wider range of
temperature by Werthamer \cite{werthamer}, and then to higher order in
the gradients of the order parameter by Tewordt \cite{tewordt} and
Eilenberger \cite{eilenberger1}.

In the approach originally due to Gor'kov and then followed by
Eilenberger, Werthamer, Tewordt, and others, the free energy is
calculated by relating its variation to the diagonal element of the
Green function introduced by Gor'kov \cite{gorkov2}.  These
authors calculated the full Green's function  in the Born series, and
set its arguments equal at the end.  Then they made an ansatz for the
free energy in the same approximation by considering all possible
terms that may enter with undetermined coefficients.  Upon the
variation of the ansatz and comparison with  the variation of the
free energy, they obtained the coefficients. This is very laborious,
and moreover nonsystematic.

Since the superconducing gap is much smaller than the Fermi energy,
it is usually safe to linearize the single-particle energy of the
normal metal in the vicinity of the Fermi surface. Once this
approximation is made, the full Gor'kov Green functions reduce to
sums of Green functions, or resolvents, of one-dimensional
differential operators \cite{waxman1}.  In the years since the
publication of the work  described above, a great deal has been
learned about the properties of  such resolvents at points where the
arguments coincide.  When the Green function concerned is that of a
Schr{\"o}dinger operator, the diagonal element of the resolvent
satisfies an equation discovered by Gelfand and Dikii
\cite{gelfand}.  This allows a fast and efficient evaluation of the
diagonal element as an asymptotic expansion in powers of the
potential and its derivatives. The terms in this expansion turn out
to be the infinite sequence of conserved hamiltonian densities for
the Korteweg-de Vries (KdV) hierarchy of integrable partial
differential equations \cite{dikey}.

For our superconductivity problem, the one-dimensional Schr{\"o}dinger
operator is replaced by a first-order matrix differential operator of
the form  studied by Andreev \cite{andreev}.  It is identical to
a one-dimensional Dirac operator with  a chiral mass term.  Once
again the diagonal element of the resolvent (or a simple modification
of this) satisfies an equation which quickly and efficiently
generates a series expansion in the coeficients of the equation and
their derivatives.  The terms of this series give the hamiltonian
densities for the  family of  integrable equations known as the  AKNS
hierarchy \cite{AKNS}, whose  simplest member is equivalent to the
non-linear Schr{\"o}dinger equation \cite{dikey2}. Perhaps
surprisingly, this generating equation has also been exploited in the
superconductivity literature.  It is nothing other than the
Eilenberger equation \cite{eilenberger2} which plays a central role in
the quasiclassical theory \cite{rainer}.

In a previous paper \cite{kosztin} we used both the Gelfand-Dikii
equation and the Eilenberger equation to generate the gradient
expansion for the free energy as far as the eighth order in the
gradients of the gap function.  Unfortunately these results are
expressed in terms of gradients of the magnitude of the gap-function
and gradients of its phase  separately. This leads to very long
expressions which are of limited utility. We were, however, able to
expand the much more compact expressions appearing in the classic
paper by Tewordt \cite{tewordt} into our form and compare them. We
found that our results differed from those of Tewordt at the fourth
order --- the highest order he had computed.  While we had confidence
in our own results  (we had derived them from two completely
different methods, using {\it Mathematica\/} to automate the labor),
we were not able to isolate the source of the discrepancy.

In this paper we report another method for deriving the gradient
expansion. The new method is not quite as efficient at generating the
series as that in \cite{kosztin}, but has the advantage that the
resulting expressions  are very compact. We evaluate all terms up to
fourth order in the gradients, and expand them out so as to express
our results in the form used by Tewordt \cite{tewordt}. We are thus
able identify what appears to be a  typographical error in his
paper.  While the correct fourth-order free energy is the principal result of
the present paper, we believe that the methods we use are interesting
in their own right, and will have application whenever one requires
corrections to the leading orders of the quasiclassical method.

In section  two we briefly review how the calculation of the free
energy of a three dimensional superconductor is reduced to computing
determinants of one-dimensional differential operators. In section
three we show how the Eilenberger equation arises as a property of the
resolvent of such differential operators. In section four we use the
Eilenberger function to derive a useful identity linking the dressing
function for the Andreev hamiltonian to the determinant we wish to
compute. In section five we show how this useful identity is related
to the ``shooting method'' for computing determinants. Then, in
section six, we use our identity to compute the gradient expansion.
Finally in section seven we compare our results with those of
reference \cite{tewordt}.

\section{From Three Dimensions to One}

In this section we will review standard material so as to establish
our notation.

Following Gor'kov \cite{gorkov2}  we treat the normal metal as a free,
highly degenerate, electron gas and the superconductivity as arising
from the BCS model potential -- an instantaneous short-range
attraction between pairs of electrons whose energy is within a narrow
shell of width $\omega_{{\rm debye}}$ about the Fermi surface. 
We also ignore paramagnetic effects.
The
partition function of such an electron gas  may be written as a
Berezin path integral
\bea
&&{\Z}=\Tr(e^{-\beta H}) \nonumber\\ 
&&=\int d[\psi]d[\psid]
\exp {-\int_0^\beta d^3x d\tau \left\{
\sum_{\alpha=1}^{2} \psid_\alpha(\partial_\tau-\frac
1{2m}\nabla^2-\mu)\psi_\alpha -
V\psid_1\psid_2\psi_2\psi_1\right\}}.\nonumber\\
\label{EQ:ch14_BCS}
\eea    
The indices $\alpha= 1,2$ refer to the two components $\uparrow,
\downarrow$, of spin.  The Grassmann-valued Fermi fields $\psi$,
$\psid$,  are to be taken  antiperiodic under  $\tau \to
\tau+\beta$.

A positive value for $V$ corresponds to an attractive interaction
between the particles. Given an attractive interaction, and a low
enough temperature, the system should be unstable  with respect to
the onset of superconductivity. To detect this instability we
introduce the ancillary  complex scalar field $\Delta$ which will
become the superconducting gap-function. We use it to decouple the
interaction by writing
\bea
&&{\Z}=\int d[\psi]d[\psid]d[\Delta]d[\Delta^*]
\exp -\int_0^\beta d^3x d\tau \left\{
\sum_{\alpha=1}^{2} \psid_\alpha(\partial_\tau-\frac
1{2m}\nabla^2-\mu)\psi_\alpha  \right.\nonumber\\
 &&\qquad\qquad \left. -\Delta^*\psi_2\psi_1-\Delta\psid_1\psid_2
+\frac {1}{V}|\Delta|^2\right \}.
\eea
The equation of motion for $\Delta$ shows us that $\Delta\equiv V
\psi_2\psi_1$.

Taking note of the anticommutativity of the  Grassmann fields,
the quadratic form in the exponent can be arranged as a
matrix
\be
(\psid_1,\psi_2)
\left(\matrix{ \partial_\tau +\left(-\frac 1{2m} \nabla^2 -\mu\right) & \Delta\cr 
              \Delta^*  &  \partial_\tau -\left(-\frac 1{2m} \nabla^2 -\mu\right)
                        \cr}\right)
\left(\matrix{\psi_1\cr\psid_2\cr}\right).
\ee
We may now integrate out the fermions, obtaining the functional
determinant of the Bogoluibov-de Gennes (BdG) operator  
\be
{\cal B}=
\left(\matrix{ \partial_\tau +\left(-\frac 1{2m} \nabla^2 -\mu\right) & \Delta\cr
                \Delta^*  &  \partial_\tau -\left(-\frac 1{2m} \nabla^2 -\mu\right)
                 \cr}\right).
\ee
The BdG operator acts on two-component Nambu spinors obeying
$\psi({\bf x}, \tau+\beta)= -\psi({\bf x},\tau)$.

Provided we can ignore quantum and thermal fluctuations of the gap
function $\Delta(x)=|\Delta|e^{i\theta}$, we can perform the integral
over $\Delta$ by simply replacing it by its saddle-point value. In
this way  the euclidean time  effective action for a BCS
superconductor is given as
\be 
\Gamma(\Delta)= -\ln\Det {\cal B} +\int d^3xdt\,\frac
{1}{V}|\Delta|^2.
\ee
In this work we will be  concerned   with the situation  where the
gap-function $\Delta$ is time-independent. In this case the
functional $\Gamma$ reduces to  $\beta{\cal F}(\Delta)$ where ${\cal
F}$ is the free energy.

It is not possible to evaluate  
$\Gamma$ exactly for an arbitrary gap-function, but   
when 
$\Delta$ is much smaller than $E_F$ and varies slowly on the scale of the  Fermi
wavelength,
then only momenta near the Fermi surface are important. In this case it is reasonable to replace 
the single-particle energies
of the normal metal by the linearized form
\be
\epsilon({\bf k}) = v_F(|{\bf k}|- k_F),
\ee
and to approximate  $k^2 d|k|$ by $k_F^2 d|k|$. 
Having done this, then with no further approximation, we may then use  the results from \cite{waxman1} to 
write
\be\label{EQ:quasiclassical}
\Gamma(\Delta)= -2\pi v_F N(0) \int\frac{d\Omega_{\n}}{4\pi}
d^2x_{\perp}
\ln\Det(\partial_\tau -iv_F \sigma_3 (\n\cdot\nabla) +
|\Delta|\sigma_1 e^{-i\sigma_3\theta}).
\ee

What is happening in  equation (\ref{EQ:quasiclassical}) requires some explanation.
Firstly $N(0)$ is the density of states at the Fermi surface 
\be 
 N(0)=
\frac {mk_F}{2\pi^2},
\ee
the symbol $\n$ denotes  a unit vector, and $v_F=k_F/m $ is the Fermi
velocity. Having chosen a  point ${\bf k}= \n\, k_F$ on the Fermi
surface, we decompose the coordinate vector ${\bf x}$ into a
part parallel to $\n$, and a part perpendicular
\be
{\bf x} = x_\parallel \n + {\bf x}_\perp.
\ee 
Then, for fixed  ${\bf x}_\perp$, 
we compute  $\ln \Det(\partial_\tau +{\cal H})$ where  
${\cal H}$ is the  one-dimensional Andreev hamiltonian \cite{andreev} 
\be
{\cal H}=-i\sigma_3v_F (\n\cdot\nabla) + |\Delta| \sigma_1 e^{-i\sigma_3
\theta}=\left[ \begin{array}{cc}  - iv_F  (\n\cdot\nabla)  & \Delta\\ 
                               \Delta^*  &+ iv_F  (\n\cdot\nabla)  
                               \end{array}\right].  
\ee
This Dirac-like  operator appears as an approximation to the full 
hamiltonian when we restrict our attention to   Nambu spinors  of the
form  $\chi(x)\exp{ik_F{\bf n}\cdot {\bf x}}$, where  $\chi(x)$ is
slowly varying.  Here the spinor $\chi$  and the gap-function $\Delta$ are
to be  regarded as functions of $x_\parallel$.  Notice that the
derivative $\n\cdot\nabla$ coincides with $\partial_{x_\parallel}$.
Having evaluated   $\ln\Det(\partial_\tau +{\cal H})$, we then
integrate the result over  the family of parallel lines  parameterized by ${\bf
x}_\perp$.  Finally the $\int {d\Omega _{\bf n}\over 4\pi }$ performs an 
average over  the directions ${\bf n}$.

If we are given an expression for  $\ln\Det(\partial_\tau +{\cal H})$  in the form
\be
\ln\Det(\partial_\tau +{\cal H})= \int_{-\infty}^{\infty} 
F(x_{\parallel}, {\bf x}_{\perp})dx_{\parallel}dt,
\ee
then  the   integrations over $x_\parallel$ and ${\bf x}_\perp$
combine  to give an integral over every point in three-dimensional
space.  To illustrate this,  consider  a time-independent $\Delta$ at
zero temperature.  To second order accuracy in the gradient expansion
the part of the free energy coming from gradients of the phase of the
order parameter is \cite{stone}
\be
\int 
\frac {v_F}{8\pi} \left( (\n\cdot \nabla) \theta\right)^2 dx_{\parallel}dt.
\ee
Using
\be
\int\frac{d\Omega_{\n}}{4\pi} n_i n_j =\frac 13 \delta_{ij},
\ee
and remembering that $k_f= mv_F$, the effective action becomes
\be
\Gamma(\Delta) = \frac {k_f^3}{3\pi^2} \int  \frac
1{8m}(\nabla\theta)^2 d^3xdt.
\label{EQ:2gamma}
\ee

This is not unreasonable. Recognising that  
\be
\frac {k_f^3}{3\pi^2} =\rho_s,
\ee
is the number-density of  the  electrons, and that the superfluid velocity
is
\be
{\bf v}_s= \frac 1{2m}\nabla\theta,
\ee
we can rewrite (\ref{EQ:2gamma}) as 
\be
\Gamma(\Delta)= \int \frac 12 \rho_s mv^2_s d^3xdt\,.
\ee
Since everything is time-independent the $t$ integral is trivial, and from   
$\Gamma(\Delta)=\beta{\cal F}(\Delta)$ we  have 
\be
{\cal F}(\Delta)= \int \frac 12 \rho_s mv^2_s d^3x.
\ee
Thus this part of the  free energy coincides with the kinetic energy of the
superflow.

\section{The Resolvent Equation}

When ${\cal H}$ is time-independent 
we have 
\be
\ln \Det(\partial_\tau +{\cal H})= 
\sum_{n=-\infty}^{\infty}\ln \Det(i\omega_n +{\cal H}),  
\ee
where $\omega_n = 2\pi(n+\frac 12)/\beta$, the fermionic Matsubara
frequences, are the eigenvalues of $\partial_\tau$. 

Our task therefore is to obtain the the Fredholm determinant $\Det
({\cal H}-E)$  involving the Andreev hamiltonian operator
\be
{\cal H}=-i\sigma_3 \partial_x + |\Delta| \sigma_1 e^{-i\sigma_3
\theta},
\label{EQ:andreevop}
\ee  
and where $E$ is a general complex number. (We have set $v_F=1$.)
Such  determinants are usually evaluated by use of  the variational formula
\be
\delta \ln \Det ({\cal H}-E)= \Tr \{({\cal H}-E)^{-1} \delta ({\cal
H}-E)\},
\label{EQ:detvary}
\ee
where   $({\cal H}-E)^{-1}$ denotes the Green function, or resolvent, 
\be
G_{\alpha\beta}(x,y,E)= \eval{\alpha,x}{({\cal H}-E)^{-1}}{\beta,y}.
\ee
The Green function  can be regarded as an infinite dimensional matrix in the
variables $x$, $y$, and as a  $2\times 2$ matrix in the Nambu spinor
space on which the $\sigma_i$ matrices act. The notation $\Tr A$
includes an integration over the $x,y$ labels as well as a
conventional trace (to be denoted by $\tr$) over the spinor indices.

Provided $E$ is not an eigenvalue of ${\cal H}$, the  Green function exists
 and obeys
\be
(-i\sigma_3 \partial_x + |\Delta|
 \sigma_1 e^{-i\sigma_3 \theta}-E)_{\alpha\gamma}G_{\gamma\beta}(x,y,E)
=\delta(x-y)\delta_{\alpha\beta}.
\label{EQ:gnfneq}
\ee
If  $\psi_\alpha^L$ and $\psi_\alpha^R$ are solutions to the equation
\be
(-i\sigma_3 \partial_x + |\Delta| \sigma_1 e^{-i\sigma_3
\theta}-E)\psi=0,
\ee
satisfying suitable boundary conditions to the left and right respectively, then
\bea
G_{\alpha\beta}(x,y,E)&=
 \frac i W\psi_\alpha^R(x)\psi_{\beta'}^L(y)(\sigma_1)_{\beta'\beta} &
{\rm for}\quad x>y\nonumber\\
&= 
\frac i W\psi_\alpha^L(x)\psi_{\beta'}^R(y)(\sigma_1)_{\beta'\beta} &{\rm for}\quad
y>x.
\eea
Here
\be
W=W(\psi^L,\psi^R) =-i (\psi^L)^T\sigma_2 \psi^R 
= \psi^R_1\psi^L_2-\psi^R_2\psi^L_1.
\ee  
The expression $W(\psi^L,\psi^R)$ is  independent of $x$  for any
$\psi^L$, $\psi^R$ that are both solutions of $({\cal H}-E)\psi=0$.
$W$ plays a role analogous to that of the Wronskian in the theory of
scalar linear differential equations. In particular $W$ vanishes
identically if and only if $\psi^L$ and $\psi^R$ are linearly
dependent. Such linear dependence signals that
$E$ is  an eigenvalue.

When making the variation    $\delta ({\cal H}-E)$  in
(\ref{EQ:detvary}) we will only  consider changes in  $E$ and
$\Delta$, so we will require $G(x,y,E)$ only at the points $x=y$. Now
strictly speaking, $G(x,x;E) $ is not well defined, since the jump
condition implied by (\ref{EQ:gnfneq}) is
$$
G(x,x+\epsilon)-G(x,x-\epsilon)=i\sigma _3.
$$
Fortunately this ambiguity does not affect 
$ \delta \ln \Det ({\cal H}-E)$. This is  because
$$
\tr \left[ \sigma _3 \delta\pmatrix{-E & \Delta  \cr
\Delta ^*  &- E }\right] =\tr \delta
\pmatrix{-E & \Delta  \cr
-\Delta ^*  & E } =0.
$$
We may therefore safely define the symbol   
$G(x,x,E)$ to mean the  average of its two limits
\cite{feinberg}\cite{shelankov}
\be
G(x,x)\stackrel{def}{=}
\lim_{\epsilon \to 0^+} \frac 12 ( G(x+\epsilon,x) +
G(x,x+\epsilon)).
\ee

With this understanding consider the  matrix-valued
function $g(x)= 2iG(x,x,E)\sigma_3$. 
In components this is  
\be
g(x)_{\alpha\beta}= 
\frac 1{W}\left( \psi^R_\alpha\psi^L_{\beta'}+ 
\psi^L_\alpha\psi^R_{\beta'}\right)(i\sigma_2)_{\beta'\beta} 
\label{EQ:eilenbergerg}.
\ee
From the equations obeyed by $\psi^R$ and $\psi^L$ we immediately see
that $g$ obeys
the equation
\be
\partial_x g= [M,g],
\label{EQ:resolventeq}
\ee
where
\be
M= -i\sigma_3 (|\Delta| \sigma_1 e^{-i\sigma_3 \theta}-E).
\ee
The reason for the appearance of $\sigma_3$ in the definition of $g$
is that $\sigma_3$ appears in the coefficient of $\partial_x$ in the Andreev
operator (\ref{EQ:andreevop}). If we pre-multiply the Andreev operator by $-i\sigma_3$, it
becomes $-\partial_x+M$, where the coefficient of $\partial_x$ is now
proportional to the identity matrix. The Green function for this modified operator
is  $G(x,y,E)\sigma_3$, and $g(x)$ is, 
up to the  factor of $2i$, the diagonal element of this Green function 
in $x$ space (it is still a matrix in Nambu  spinor space).

The equation (\ref{EQ:resolventeq}) can also be written as a
commutator
\be
[-\partial_x+M,g]=0,
\ee
where we regard $g(x)$ as an operator acting on the space of vector valued functions
by pointwise multiplication.

In the superconductivity literature (\ref{EQ:resolventeq}) is usually
called  the {\it Eilenberger equation\/} and $g$ the {\it Eilenberger
function\/} \cite{eilenberger2}. It lies at the heart of the
quasiclassical theory of superconductors \cite{rainer}.  Equation
(\ref{EQ:resolventeq}) is also the analogue, for matrix first-order
differential operators, of the Gelfand-Dikii equation
\cite{gelfand}\cite{kosztin}.  Like the Gelfand-Dikii equation, it
has extensive applications in the theory of integrable dynamical
systems \cite{dikey2}. Because we wish to stress that
(\ref{EQ:resolventeq})  is an {\it exact\/} property of the diagonal
element of the resolvent of any first-order matrix differential
operator of the form $-\partial_x + M$ (rather than an {\it
approximate\/} property of the partially integrated Gor'kov Green
function) we will often refer to it as the {\it resolvent
equation\/}.

Using the definition of $W$, it is clear that the  particular solution
of the resolvent equation given by equation (\ref{EQ:eilenbergerg})
satisfies the condition  $\tr g=0$. Similar algebra shows that 
\be
g^2 \psi^L=\psi^L,\qquad g^2 \psi^R=\psi^R.
\ee 
Since the two solutions are linearly independent and span the space
of Nambu spinors at each $x$, this implies $g^2= {\bf 1}$. This
condition is sufficient to determine the solution\footnote{The
normalization $g^2={\bf 1}$ is specific to the
resolvent of  the Andreev operator on an infinite line, or on a
finite interval with self-adjoint boundary conditions at each end.  When periodic
boundary conditions are imposed on an interval of length $L$, the
normalization becomes $g^2=-{\bf 1}\cot^2 kL/2$, where $k$ is the Bloch
momentum corresponding to $E$ in the periodically extended problem.}.

Given  $g$ as a function of $E$, it is possible to recover 
$\ln\Det({\cal H}-E)$. We simply observe that
\be  
\frac {\partial}{\partial E}\ln\Det({\cal H}-E)= 
-\int dx\, \frac{1}{2i}\tr
(g\sigma_3),
\ee 
and integrate  with respect to $E$.

 To obtain a gradient expansion
of $g$ we introduce a parameter $z$ (which will ultimately be set
equal to $1$) and rewrite (\ref{EQ:resolventeq}) as
 \be
\partial_x g= [zM,g].
\ee 
If we  seek solutions in the form
\be
g= \sum_{n=0}^{\infty}g_n z^{-n},
\ee
we obtain the deceptively simple recurrence equation 
\be
\partial_x g_{n-1}= [M,g_n].
\label{EQ:recur}
\ee 
At first sight this determines $g_n$ in terms of a derivative of
$g_{n-1}$, so the $n$-th term contains $n$ gradients of the gap parameter.

Unfortunately (\ref{EQ:recur}) is not quite as innocent  as it looks.
The $n$-th recurrence relation
only immediately determines $g_n$ up to the addition of terms that
commute with $M$. All is not lost however. Because $M$ is a traceless
$2\times 2$ matrix, the only undetermined  term  is in fact proportional to
$M$ --- and it  is possible to find this term   from the  next
equation in the series
\be
\partial_x g_{n}= [M,g_{n+1}].
\ee
Taking  a trace shows that $\tr{M\partial_x g_{n}}=0$, and this
contains the information we need.  Using this method we have
evaluated the gradient expansion up to eighth order in the gradients
\cite{kosztin}.  Because extracting the term $\propto M$ requires an
explicit parameterization of the space of $2\times2$ matrices, the
final expressions are given in terms of gradients of $|\Delta|$ and
$\nabla \theta$ separately.  It is not at all obvious how to assemble
these very lengthy expressions  into more compact polynomials in
derivatives of $\Delta$ itself. In the following sections we will
descibe an alternative approach that does not require dissecting $g$
into parts parallel and perpendicular to $M$, and indeed does not
require us to find $g$ directly at all.

\section{A Useful Identity}

In what follows we will concentrate on evaluating the determinant in
the case where ${\cal H}$ is defined on the infinite real line ${\bf
R}$.  For convenience we will assume that all $x$ dependence of the
gap function $\Delta(x)$ takes place in a finite region $\Omega_0$,
and that to the left and right of $\Omega_0$ lie regions $\Omega_{L}$
and $\Omega_{R}$, respectively, in which  $\Delta$ takes constant
values $\Delta_{L}$ and $\Delta_{R}$.

What we do next is motivated by the discussion of first-order matrix
differential operators in \cite{dikey2}.  We observe that because
$g$ is a $ 2\times 2$ traceless matrix, it has distinct eigenvalues
and is therefore  diagonalizable.  Because $g$ also obeys the
equation  $g^2={\bf 1}$, we see that these eigenvalues are $\pm 1$.
There therefore exists a $2\times 2$ matrix $\phi(x)$ such that
$g(x)$ can be written $ g(x)= \phi(x) \sigma_3 \phi^{-1}(x)$.

As mentioned earlier   the  resolvent equation
can be written\footnote{We hope that
it will be clear from the context when a symbol such as $\partial_x g$
refers to the derivative $g'$, and when $\partial_x g$ is an operator
acting on functions according to $(\partial_x g)\varphi =
g'\varphi+g\varphi'$.} 
\be 
[-\partial_x +M, g]=0.
\label{EQ:resolventop}
\ee

Substituting $g= \phi\sigma_3\phi^{-1}$ into (\ref{EQ:resolventop})
gives \be \phi[ \phi^{-1}(-\partial_x +M)\phi , \sigma_3]\phi^{-1}=0,
\ee and this implies that \be [ \phi^{-1}(-\partial_x +M)\phi ,
\sigma_3]=0.  \ee Consequently, if we define a matrix $\Lambda$ by
\be \phi^{-1}(-\partial_x +M)\phi =-\partial_x +\Lambda,
\label{EQ:Lambdadef} \ee then $\Lambda\equiv \phi^{-1} M\phi -
\phi^{-1}\partial_x \phi$ commutes with $\sigma_3$ and so must be
{\it diagonal\/}.  The matrix $\phi(x)$ is sometimes called a {\it
dressing function\/}.

We now derive a useful identity that connects  $\Lambda$ with the
determinant we are wish  to compute. We claim that
\bea
\tr (g\delta M ) &=& \delta\tr (\Lambda \sigma_3)- \frac 12 \partial_x
\tr(\phi \sigma_3 \delta \phi^{-1} -\delta \phi \sigma_3 \phi^{-1})\nonumber\\
                 &\equiv& \delta L +\partial_x\alpha.
\label{EQ:simonseq}
\eea
(Here and elsewhere $\delta \phi^{-1}$ denotes $\delta(\phi^{-1})$ and not
$(\delta \phi)^{-1}$.)

We call the two quantities $L$ and $\alpha $ because of the analogy
with classical mechanics. In mechanics we have
\begin{equation}
\label{classmech}
\sum\limits_i \left(-m_i \ddot q_i - {\partial V \over \partial q_i}
\right) \delta q_i =
\delta \sum\limits _i \left( {1\over 2} m_i \dot q _i ^2 -V\right) +
\partial _t \left( -\sum\limits _i m_i \dot q_i \delta q_i \right).
\end{equation}
Here the first term on the right-hand side is the variation of
lagrangian, while the term in the total derivative generates the
symplectic form which  determines the hamiltonian structure.  The
left-hand side contains the functional derivative of the action, and
gives the equations of motion via D'Alembert's principle.  Dikii
\cite{dikey} exploits this analogy when discussing the stationary
equations of the integrable hierarchies.

Notice that
the second  term on the right hand side of 
(\ref{EQ:simonseq}) can equally well be 
written as 
$ -\partial_x\tr(\phi \sigma_3 \delta(\phi^{-1}) )$ 
or as $\partial_x \tr(\delta \phi \sigma_3 \phi^{-1})$.
This  is easily seen by using $\phi \phi^{-1}={\bf 1}$.

To establish (\ref{EQ:simonseq}) we rewrite (\ref{EQ:Lambdadef}) as
\bea
-\partial_x\phi +M\phi &=& \phi \Lambda\\
\partial_x \phi^{-1} +\phi^{-1} M &=& \Lambda \phi^{-1}
\eea
From these we have
\bea
&&A\stackrel{def}{=}-\partial_x \delta\phi
  + \delta M \phi +M \delta \phi - \delta \phi \Lambda
- \phi \delta \Lambda =0\\
&&B\stackrel{def}{=}\partial_x \delta \phi^{-1}
 + \phi^{-1} \delta M + \delta \phi^{-1} M - \Lambda \delta
\phi^{-1} - \delta \Lambda \phi^{-1}=0
\eea
Thus
\be
0= \tr (A\sigma_3 \phi^{-1} +\phi \sigma_3 B).
\ee
Expanding this out, using the cyclicity of the trace, the equations
$\phi^{-1} M = \Lambda \phi^{-1} -\partial_x\phi^{-1}$ and  $ M\phi =
\phi \Lambda+\partial_x\phi$, together with $g= \phi
\sigma_3\phi^{-1}$, we find (\ref{EQ:simonseq}).

To see the utility of (\ref{EQ:simonseq}) observe that 
from $g=2iG\sigma_3$ and $\delta M= -i\sigma_3 \delta({\cal H}-E)$ we have
\be
\Tr (g\delta M )= \int_{-\infty}^{\infty} dx\, \tr(g(x)\delta M(x) )= 
2\Tr(G\delta {\cal H})=2 \delta \ln \Det ({\cal H}-E).
\ee
Thus, if it is legitimate to ignore the boundary terms coming from  the total
derivative, we have
\be  
 \ln \Det ({\cal H}-E)=
 \int_{-\infty}^{\infty}  \frac 12 \tr (\Lambda \sigma_3)dx +
\hbox{\rm constant.}
\label{EQ:dettolambda}
\ee
Here the ``constant'' is a quantity unaffected by local changes in
$\Delta(x)$. It will, however, depend on the asymptotic gap functions
$\Delta_{L,R}$.

In general the boundary terms cannot be ignored, but  it is possible
to {\it choose\/} $\phi$ so that they are zero.  We have this
freedom because requiring that $g(x)= \phi(x) \sigma_3\phi^{-1}(x)$
does not uniquely determine $\phi$. Indeed the replacement $\phi \to
\phi \chi$ with $\chi(x)$ any invertable diagonal matrix leaves
$g(x)$ unchanged. Such a ``gauge transformation'' does however alter
$\Lambda$. Under the above substitution we have
\be 
\Lambda \to \Lambda - \chi^{-1}\partial_x\chi.
\ee
By making such a gauge transformation we can transfer any  of the
constributions to $\tr (g\delta M )$ from  $\Lambda$ to the total
derivative term, or vice versa. Two extreme choices  come immediately
to mind.

\begin{enumerate}

\item Select $\phi$ so that $\Lambda\equiv 0$. In this case all the
contributions to  the determinant come from the
boundary term.

\item Select $\phi$ so that the boundary terms  constribute
 nothing to the determinant.

\end{enumerate}

The latter choice is possible because the Green function $G(x,y,E)$
is insensitive to the values of $\Delta$ at large distance  from $x$
and $y$. We have assumed that there are asymptotic regions
$\Omega_{L,R}$ where $\Delta(x)$ becomes independent of $x$. Once $x$
is well inside either of these regions then $g(x)$ will settle down
to a constant value  depending only on the asymptotic values
$\Delta_{L,R}$.  We  can therefore select a $\phi$ depending only on
$\Delta_{L,R}$ to diagonalize this constant matrix.  Once we have
done this  then $\phi(\pm\infty)$ will be unaltered by variations of
$\Delta(x)$ taking place in  $\Omega_0$. The determinant is then
obtained entirely from  $\Lambda$.  This is the route we will take to
compute the gradient expansion.

\section{Connection with ``Shooting Method''}

Before we derive the gradient expansion, we should point out that our
first choice of $\phi$, the one  leading to $\Lambda=0$, is
equivalent to the so-called  ``shooting method'' for evaluating the
determinant \cite{waxman2}. This asserts that, in a suitably
interpreted sense, the determinant is proportional to the  inverse of
the transmission coefficient for scattering off the spatially varying
$\Delta(x)$. To see this, observe that if  we set $\Lambda=0$, so that
\be
\phi^{-1}(-\partial_x +M)\phi = -\partial_x +0,
\ee
then we are requiring the columns of $\phi$ to be solutions of
\be
(-\partial_x +M)\psi=0.
\ee
Equivalently
\be 
(-i\sigma_3 \partial_x + |\Delta| \sigma_1 e^{-i\sigma_3
\theta}-E)\psi=0.
\ee
A candidate for $\phi$ is therefore
\be
\phi= \left[ \begin{array}{cc} \psi_1^R & \psi_1^L\\ 
                               \psi_2^R & \psi_2^L\end{array}\right],
\ee
where the colums are composed of the $\psi^L$ and $\psi^R$ solutions
used in constructing the Green function. The inverse is
\be
\phi^{-1} = \frac 1 W \left[ \begin{array}{cc} \psi_2^L & -\psi_1^L\\ 
                               -\psi_2^R & \psi_1^R\end{array}\right]
\ee
where, as before, $W= W(\psi^L,\psi^R)=\psi_1^R\psi_2^L-\psi_2^R\psi_1^L$.

A short computation shows
that $\phi\sigma_3\phi^{-1}=  g(x)$, so  we have
guessed correctly.

Now we write down $\psi^L$ and $\psi^R$ in the asymptotic regions,
which is the only place where we will need them. Let us assume that
$E$ is real and that $E^2>|\Delta_{L}|^2$, $|\Delta_{R}|^2$, 
so we have  scattering solutions.

When $\Delta=|\Delta|e^{i\theta}$ is constant,  a  plane-wave solution to
\be
(-i\sigma_3 \partial_x + |\Delta| \sigma_1 e^{-i\sigma_3
\theta})\psi=E\psi
\ee
is given by
\be
\psi= u(k,E,\Delta)e^{ikx}= \left[ \begin{array}{c} E+k\\ 
                              \Delta\end{array}\right] e^{ikx}.
\ee
Here $E^2=k^2+|\Delta|^2$.

We now introduce  transmission and reflection coefficients, $t_{R,L}$ and $r_{R,L}$
and define
\bea
\psi^R(x)&= u(k,E, \Delta_L)e^{ikx} + r_L u(-k,E, \Delta_L)e^{-ikx}&
\quad x\in \Omega_L\nonumber\\
&= t_L u(k,E, \Delta_R)e^{ik'x} &\quad x\in \Omega_R
\eea
\bea
 \psi^L(x) &= u(-k,E, \Delta_R)e^{-ik'x}
 + r_R u(-k',E, \Delta_L)e^{ik'x}&\quad x\in
          \Omega_R\nonumber\\
           &= t_R u(k,E, \Delta_L)e^{-ikx} & \quad x\in \Omega_L.
\eea
The apparently  perverse appearance of the subscript $L$  on the
reflection and transmission coeficients in $\psi^R$  (and {\it
mutatis mutandis\/} in $\psi^L$) is supposed to indictate that the
incoming wave was incident from the left (right). The wavenumbers $k$
and $k'$ are not necessarily equal since we do not assume that
$|\Delta_L|=|\Delta_R|$.

Substituting  these expressions into $W(\psi^L,\psi^R)$ we
find that 
\bea
W(\psi^L(x),\psi^R(x))= -i (\psi^L)^T\sigma_2\psi^R
                &= -2k\Delta_L t_R &\quad x\in \Omega_L\nonumber\\
                &= -2k'\Delta_R t_L &\quad x\in \Omega_R.
\eea
Since $W$ is $x$ independent, we must have 
\be
k\Delta_L t_R=k'\Delta_R t_L. 
\ee
This  reduces
to $t_L=t_R$ in the particular case that $\Delta_L=\Delta_R$.

In order to use $\psi^R$ and $\psi^L$ in the Green function we
require that  $k$, $k'$, have a  positive imaginary part so that
$\psi_R$ tends to zero as $x\to +\infty$, and similarly $\psi_L$
tends to zero as $x\to -\infty$. If we assume that both the real part
of $E$ and the real part of $k$ are positive  (so that $\psi^R$ and
$\psi^L$ correspond to a real scattering process, and $G$ to outgoing
waves) then this requires the addition of  positive imaginary part to
$E$.  If we wish to evaluate $G$, and from this  $\Det({\cal H}-E)$,
for $E$ below the positive real axis then we must replace $k$ by $-k$
in the above wavefunctions, so that the resulting negative imaginary
part of $k$ still leads to convergence. As expected, this means that
both $G$ and $\Det({\cal H}-E)$ have a branch cut discontinuity
across the real $E$ axis whenever asymptotic plane-wave solutions are
possible.

Now we use these functions to evaluate $\tr(\delta \phi \sigma_3
\phi^{-1})$. Near $x=-\infty$ we have $\psi^R \to  u(k,E,\Delta_L)
e^{ikx}$ which is large, while $\psi^L= t_Ru(k,E,\Delta_L) e^{-ikx}$
is small. The only expressions occuring in $\tr(\delta \phi \sigma_3
\phi^{-1})$ involve their product which is finite. Varying $\Delta$
in $\Omega_0$ changes  $t_R$, while leaving $u(k,E,\Delta_L)$ {\it
etc.\/} unaltered.  We find that
\be
\left.\tr(\delta \phi \sigma_3
\phi^{-1})\right|_{x=-\infty}= \frac {\delta t_R}{t_R}.
\ee
Similarly 
\be
\left.\tr(\delta \phi \sigma_3
\phi^{-1})\right|_{x=+\infty}= - \frac  {\delta t_L}{t_L}.
\ee

Inserting  these results into (\ref{EQ:simonseq}) leads to
\be 
\ln \Det({\cal H}-E) = -\ln t + {\rm constant}.
\ee

It does not matter whether we use $t_L$ or $t_R$ in this formula
because the logarithms differ by terms involving $E$ and
$\Delta_{R,L}$, and these can be included in the constant.  The
constants cancel when we consider ratios of determinants  of
operators with the same asymptotic $\Delta$'s. In other words, if ${\cal
H}$ and ${\cal H}^{(0)}$ are two hamiltonians with the same asymptotic
behaviour, then
\be 
\frac {\Det({\cal H}-E)}{\Det({\cal H}^{(0)}-E)}
= \frac {t^{(0)}_L(E)}{t_L(E)}
=\frac {t^{(0)}_R(E)}{t_R(E)}. 
\ee

Notice that the (analytic continuation of) $t$ and $r$ become
infinite as $E$ approaches the energy of a  bound state.  The
wavefunctions  $\psi^R$ and $\psi^L$ become proportional to one
another, and  decay exponentially as $|x|\to \infty$. The resultant
vanishing of the determinant is consistent with the interpretation of
the ratio of the Fredholm  determinants as a regularized version of
\be
\prod_{n} \frac{(E_n-E)}{(E_n^{(0)}-E)},
\ee 
where the $E_n$ are the eigenvalues of ${\cal H}$, and the
$E_n^{(0)}$ the eigenvalues of ${\cal H}^{(0)}$.  For $E$ in the
continuous spectrum the zeros and poles coming from the eigenvalues
of ${\cal H}$ and ${\cal H}^{(0)}$ merge to form the branch-cut noted
above.

\section{The Gradient Expansion}

To find the recurrence relation for $\phi$ we multiply 
 the  matrix $M$ by $z$ as before.
At the end of the calculation,
we may set $z=1$. Thus again we are looking for a solution of
\begin{equation}
\label{zeilen}
[-\partial + zM, g]=0
\end{equation}
as a power series in $1/z$. To find the free energy expansion,
we do not need $g$ itself but only $\phi$ and $\Lambda $ that satisfy
\begin{equation}
\label{zlambda}
zM\phi = \phi ' + \phi z \Lambda,
\end{equation}
and are given as power series in $1/z$.

For the homogeneous superconductor, only the zeroth order term in
$\Lambda$ will be non-zero. We find it from the zeroth order
expression for $g$. Since $g$ is traceless, squares to the identity,
and commutes with $M$, we must have
\be
g_0= \frac 1{\zeta} M,
\label{EQ:g_0}
\ee
where $\zeta$ is defined by $M^2=\zeta^2 1$. Since
\be
M=\left(\matrix{\omega_m & -i\Delta\cr
                i\Delta^* & -\omega_m\cr}\right)
\ee
(we have now replaced $-E$ by the Matsubara frequencies)
we have
\be
\zeta =\sqrt{\omega_m^2+|\Delta(x)|^2}.
\ee     
From (\ref{EQ:simonseq}) and (\ref{EQ:g_0}) we find
\be
\delta \tr(\sigma_3 \Lambda_0)= \tr (g_0 \delta M)= 2 \delta \zeta,
\ee
so 
\begin{equation}
\tr (\Lambda _0 \sigma_3) = 2 \zeta,
\end{equation}
and
\be
\Lambda_0= \sigma_3 \zeta.
\ee 

Hence, we look for $\Lambda $ in
the form
\begin{equation}
\label{zlamb}
\Lambda = \zeta \sigma_3 +{\Lambda _1 \over z} + {\Lambda _2 \over z^2} + \dots
\end{equation}
Comparing coefficients of $z^1$ in (\ref{zlambda}), we obtain the equation
for the 0-th order in $\phi $:
\begin{equation}
\label{diagM}
M\phi _0 = \phi _0 \zeta \sigma_3
\end{equation}
To obtain a recurrence relation for higher-order terms in $\phi $, it 
turns out that $\phi _0$ has to be factored out from the expansion, i.e.,
we look for $\phi $ in the form
\begin{equation}
\label{zphi}
\phi = \phi _0\left( \1 +{\phi _1 \over z}+ {\phi _2 \over z^2} + 
\dots \right).
\end{equation}
Now we can go on to calculate the first-order terms in (\ref{zlamb})
and (\ref{zphi}) by comparing the coefficients in (\ref{zlambda}):
\begin{equation}
M\phi _0 \phi _1 = \phi _0 ' + \phi _0 \Lambda _1 +
\phi _0 \phi _1 \zeta \sigma_3
\end{equation}
Multiplying by $\phi _0 ^{-1}$ on the left and using (\ref{diagM}),
we obtain
\begin{equation}
\label{m1}
\zeta [\sigma_3,\phi _1]-\Lambda _1 = \phi _0 ^{-1} \phi _0'.
\end{equation}
In general, comparing the coefficients of $z^{-n}$ in (\ref{zlambda}),
we get the $n-$th order relation
\begin{equation}
\label{mrectransf}
\zeta [\sigma_3,\phi _{n+1}]-\Lambda_{n+1}= \phi _n' + 
\phi _0 ^{-1} \phi _0 ' \phi _n + \phi _n \Lambda _1 + 
\phi _{n-1} \Lambda _2 + \dots + \phi _1 \Lambda _n.
\end{equation}
The $(n+1)$th terms in the expansions of $\phi$ and $\Lambda$ are
thus given by all the previous terms. Moreover the first term on the
left-hand side of (\ref{mrectransf}) is perpendicular to $\sigma_3$,
whereas $\Lambda $ contains a multiple of $\sigma_3$ and a multiple
of $\1 $, since it commutes with $\sigma_3$.  In other words, the
first term on the left-hand side contains $\sigma _1$ and $\sigma
_2$, whereas the second one contains $\sigma _3 $ and $\1 $. Hence,
we obtain the $(n+1)$-th terms in the expansion of $\phi $ and
$\Lambda $ by calculating the right-hand side of (\ref{mrectransf})
(in which everything is known from the preceding steps in the
recursion), and splitting it up into the transverse and longitudinal
parts relative to $\sigma_3$.

There is a minor hitch: (\ref{mrectransf}) does not determine the
part of $\phi _{n+1}$ that commutes with $\sigma_3$. Adding a term to
$\phi _{n+1}$ that commutes with $\sigma_3$ {\it does \/} change the
right-hand side of all subsequent recurrence relations. Such a term
may, however, always by removed by changing our choice of the  $\phi$
that diagonalises $g$.  This will change  the free energy density by
a total derivative of terms local in the gap. It will not affect the
total free energy.

We may therefore  proceed to find $\phi$ and $\Lambda$. Unfortunately
this  will lead to the same lengthy expressions we found in
\cite{kosztin} by evaluating the recurrence relation for $g$.  The
problem is that  there is no simple way to write the  matrix
$\phi_0^{-1}{\phi_0}'$ in terms of $M$ itself. We will therefore have
to choose an explicit basis for the matrices and work with
components. 

After a lot of thought we realized that is far more
convenient to replace the arbitrarily chosen matrix $\sigma_3$ in the
decomposition $g= \phi \sigma_3 \phi^{-1}$ by the local matrix $M$.
To do this  we transform
\begin{eqnarray}
\phi \rightarrow \Phi & \equiv & \phi \phi _0 ^{-1} =
\1 + {\phi_0 \phi_1 \phi _0 ^{-1} \over z} + 
{\phi_0 \phi_2 \phi _0 ^{-1} \over z^2 } + \dots \nonumber \\
& \equiv & \1 + {\Phi _1 \over z} + {\Phi _2 \over z^2} + \dots \nonumber \\
\Lambda \rightarrow \l & \equiv & \phi _0 \Lambda  \phi _0^{-1}=
M + {\phi _0 \Lambda _1 \phi _0^{-1}\over z}+
{\phi _0 \Lambda _2 \phi _0^{-1}\over z^2} + \dots \nonumber \\
& \equiv & M + {\l _1\over z} + {\l _2 \over z^2 } + \dots
\end{eqnarray}
Then $ L\equiv \tr (\Lambda \sigma_3) = \tr (\l M/\zeta )$. By using
$$
\phi _0 \phi _n' \phi _0^{-1} = \Phi _n' - 
\phi _0 ' \phi _0 ^{-1} \Phi _n +
\Phi _n \phi _0 ' \phi _0 ^{-1},
$$
we can rewrite (\ref{mrectransf}) as
\begin{equation}
[M,\Phi _{n+1}]- \l _{n+1} = \Phi _n' + \Phi _n \phi _0 ' \phi _0 ^{-1}+
\Phi _n \l _1 + \dots + \Phi _n \l _1 .
\end{equation}

The advantage of this expression is that the two   factors of $\Phi
_n$ in the second and third terms on the right hand side  are both on
the left.  This enables us to get rid of the awkward expression $\phi
_0 ' \phi _0 ^{-1}$ by factoring  out $\Phi _n$  and using the
recurrence relation for $n=0$
$$
[M,\Phi _1]-\l _1 = \phi _0 ' \phi _0 ^{-1}.
$$
We obtain
\begin{equation}
\label{Mrecurrence}
[M,\Phi _{n+1}]- \l _{n+1} = \Phi _n' + \Phi _n [M, \Phi _1]+
\Phi _{n-1} \l _2 + \dots +  \Phi _n \l _1 .
\end{equation}
We need, however, to start the recurrence  by obtaining  $\Phi _1$ in
some  way.  For this we can  use the Eilenberger equation
(\ref{zeilen}). From (\ref{EQ:g_0}) and (\ref{diagM}), we get the
expansion of $g$ in $1/z$
\begin{eqnarray*}
g&=& \phi \sigma_3 \phi ^{-1} = \Phi {M\over \zeta} \Phi ^{-1} =
{M\over \zeta} + {1\over z} \left[\Phi _1, {M\over \zeta}\right] +\dots \\
& \equiv & g_0 +{g_1\over z} + \dots
\end{eqnarray*}
The coefficient of $z^0$ in (\ref{zeilen}) gives
$$
g_0'=[M,g_1],
$$
that is,
$$
\left( {M\over \zeta } \right) ' = \left[ M, \left[ \Phi _1,
{M\over \zeta } \right] \right].
$$
The double commutator of
 matrices $A,B,C$ can be written as
\begin{equation}
\label{doublecomm}
[A,[B,C]]=\{\{A,B\},C\} - \{B,\{A,C\}\}.
\end{equation}
Here the braces denote an {\it anti-commutator\/}.
For three traceless $2\times 2 $ matrices this reduces to a form of
the vector triple-product identity
$$
[A,[B,C]]=2\tr (AB) C - 2\tr (AC) B.
$$

We now choose $\Phi _1$ with no longitudinal components, i.e.
$\tr \Phi _1 = \tr (\Phi _1 M) =0$. Using 
$\tr M^2 = 2 \zeta ^2 $, we get
\begin{equation}
\label{Phi1}
\Phi _1 = -{1\over 4} {1\over \zeta } 
\left( M\over \zeta \right)'.
\end{equation}
We see the natural appearance of $M/\zeta$ as well as of the 
derivative multiplied by $1/\zeta$. 
Hence, it is natural to divide (\ref{Mrecurrence}) by $\zeta $, and to
define
\begin{eqnarray*}
N & \equiv & {M\over \zeta }, \\
\L & \equiv & {\l \over \zeta} =
N+ {1\over z} {\l _1 \over \zeta } + {1\over z^2} {\l _2 \over \zeta }
+ \dots \\
& \equiv & N + {\L _1 \over z} + {\L _2 \over z^2 } + \dots,
\end{eqnarray*}
and
\begin{equation}
\label{D}
D \equiv {1\over \zeta } \partial.
\end{equation}
The symbol $D$ still behaves as a derivative in the sense that it is
linear, and obeys the Leibnitz rule
$$
D(AB) = (DA)B + ADB.
$$
In terms of these new variables, we can rewrite (\ref{Mrecurrence}) as
\begin{equation}
\label{Nrecurrence}
[N,\Phi _{n+1}]-\L_{n+1}=D\Phi _n + \Phi _n[N,\Phi _1] +
\Phi _{n-1} \L_2 + \dots + \Phi _1 \L_n,
\end{equation}
and
$$
 L= \zeta \tr (\L N).
$$

The formula (\ref{Nrecurrence}) leaves undetermined the part of 
$\Phi _{n+1}$ that commutes with $N$ (just as (\ref{mrectransf})
did not determine the part of $\phi _{n+1}$ that commuted with $\sigma_3$).
Again it is convenient to choose the $\Phi $'s purely transverse,
i.e.,
$$
\tr \Phi_k = \tr (\Phi_k N) = 0, \quad \mbox{for all} \ k\geq 1.
$$
With this choice  we  get a recurrence relation for the $\Phi $'s only,
and a very simple expression of $L$ in terms of the $\Phi $'s.

To see how this comes about,
let us think of $N$ as being $\sigma _3$ (this can always be achieved
at a given point by a global rotation which does not change $L$),
then the $\Phi $'s contain $\sigma _1$ and $\sigma _2$, whereas the
$\L $'s contain $\1 $ and $\sigma _3$. Hence, among the terms in
(\ref{Nrecurrence}), $\L _{n+1}$ and $\Phi _n [N,\Phi _1] $ contain
$\1 $ and $\sigma _3 $ only, $[N,\Phi _{n+1}]$ and all the products of
the type $\Phi _j \L _k $ contain $\sigma _1$ and $\sigma _2$ only.
Finally $D\Phi _n$ potentially contains all four matrices. We can
separate the transverse and the longitudinal part by taking the commutator
and anticommutator of $N$ with both sides of (\ref{Nrecurrence}). From
the commutator we get
\begin{equation}
\label{com}
[N,[N,\Phi _{n+1}]]=[N,D\Phi _n] + 
[N,\Phi _{n-1} \L _2 + \dots + \Phi _1 \L _n].
\end{equation}
Using  the double commutator formula (\ref{doublecomm}) and its special
case for traceless matrices, this leads to
\begin{equation}
[N,[N,\Phi _{n+1}]]=
2\tr (N^2) \Phi _{n+1} - 2\tr (N \Phi _{n+1}) N = 4\Phi _{n+1},
\end{equation}
where the second term in the middle expression is zero due to transversality.
For the last term on the right-hand side of (\ref{com}) we use the formula
$$
[A,BC]=\{A,B\}C-B\{A,C\} 
$$
to get
\begin{equation}
\label{NPhiLam}
[N,\Phi _k \L _j] = \{N,\Phi _k\} \L _j - \Phi _k \{N,\L _j\}.
\end{equation}
Since $\{\sigma _3, \sigma _{1,2}\} =0$, the first term on the right-hand
side of (\ref{NPhiLam}) is zero. So we can rewrite (\ref{com}) as
\begin{equation}
\label{minus}
4 \Phi _{n+1} = [N,D\Phi _n] - \Phi _{n-1} \{N,\L _2\} - \dots -
\Phi _1 \{N,\L _n\} .
\end{equation} 
We see again the usefulness of introducing $N, \L,$ and $D$.

Notice the appearance of the anticommutators $\{N,\L _j\}$ on the
right-hand side of (\ref{minus}). Due to transversality of the
$\Phi$'s, the $\L $'s themselves are eliminated. We can relate these
anticommutators to the $\Phi $'s by taking the anticommutator of $N$
with both sides of (\ref{Nrecurrence}).
\begin{eqnarray}
\label{anticom}
-\{N,\L _{n+1}\} & = & \{N,D\Phi _n \} +
\left\{ N, \Phi _n [N,\Phi _1] \right\} = \nonumber \\
& = & D\left( \{N,\Phi _n\} \right) - \{DN, \Phi _n\} -
{1\over 4} \left\{N,\Phi _n [N,DN] \right\}.
\end{eqnarray}
The first term on the right-hand side of (\ref{anticom}) is zero just
as in (\ref{NPhiLam}). In the last term, we use
$$
0=D\1 = D(N^2) = N\cdot DN + (DN) N
$$
to see that 
$$
[N,DN] = 2 N\cdot DN.
$$
Then we use the formula
$$
\{A,BC\}=\{A,B\} C - B[A,C]
$$
to get
\begin{equation}
\label{lastterm}
-{1\over 4} \{N,\Phi _n 2N \cdot DN\} = - {1\over 2} \{N,\Phi _n\} N\cdot DN +
{1\over 2} \Phi _n [N,N\cdot DN].
\end{equation}
The first term on the right-hand side of (\ref{lastterm}) is again zero;
for the second one, we need to evaluate
$$
[N,NDN]=N^2DN - N(DN)N = DN + N\cdot N\cdot DN = 2 DN,
$$
where we used again the anticommutation of $N$ and $DN$.
So we can rewrite (\ref{anticom}) as
\begin{eqnarray}
\label{plus}
-\{N,\L _{n+1}\} & = & -\{DN, \Phi _n\} + \Phi _n\cdot  DN = \nonumber \\
& = & - (DN)\cdot  \Phi _n.
\end{eqnarray}
We introduce the symbol 
\begin{equation}
{\cal L}_k \equiv \{N,\L _k \} ,
\end{equation}
and substitute from (\ref{plus}) to (\ref{minus}) to get a recurrence
relation for the $\Phi $'s only:
\begin{equation}
\label{Phirecurrence}
4\Phi _{n+1} = [N,D\Phi _n] - \Phi _{n-1} (DN) \Phi _1 - \dots
- \Phi _1 (DN) \Phi _{n_1}.
\end{equation}
We can then simply calculate the terms in the expansion of
${\cal L}\equiv \{N,\L \} $ as 
\begin{equation}
\label{Lrel}
{\cal L}_n = (DN)\cdot  \Phi _{n-1}.
\end{equation}
This gives us the desired lagrangian as
\begin{equation}
\label{Llag}
L = {\zeta \over 2} \tr {\cal L}.
\end{equation}

The last three equations provide all the  machinary needed to
compute  the first few terms in the expansion of $L$. We start with
$$
\Phi _1 = -{1\over 4} DN
$$
from (\ref{Phi1}). Then
\begin{eqnarray*}
4\Phi _2 & = & [N,D\Phi _1] =\\
& = & -{1\over 4} [N,D^2 N],
\end{eqnarray*}
so
$$
\Phi _2 = -{1\over 16} [N,D^2 N].
$$
In the third order, we have
\begin{eqnarray*}
4 \Phi _3 & = & [N, D\Phi _2] - \Phi _1 (DN) \Phi _1 =\\
& = & -{1\over 16} \left[N, D[N,D^2 N] \right] - {1\over 16} (DN)^3 = \\
& = & -{1\over 16} \left[N,[DN, D^2 N] \right]
-{1\over 16} \left[N,[N,D^3 N] \right] - {1\over 16} (DN)^3
\end{eqnarray*}
We now use the formula (\ref{doublecomm}) for the double commutator and get
\begin{eqnarray*}
\Phi _3  = & - & {1\over 64} \left\{ \{N,DN\} , D^2N \right\} +
{1\over 64} \left\{ DN,\{N,D^2 N\} \right\} - \\
& - & {1\over 64} \left\{ \{N,N\}, D^3N \right\} +
{1\over 64} \left\{N,\{N,D^3N\} \right\} -{1\over 64} (DN)^3.
\end{eqnarray*}

These expresions can be simplified.  
We  begin by evaluating the  anticommutators
\begin{eqnarray*}
\{N,N\} & = & 2 N^2 = 2\1 \\
\{N,DN\} & = & 0 \\
\{N,D^2N\} & = & D\left( \{N,DN\} \right) -\{DN,DN\} = - 2(DN)^2 \\
\{N,D^3N\} & = & D\left( \{N,D^2N\} \right) -\{DN,D^2N\} =
-2D\left( (DN)^2\right)  - D\left( (DN)^2 \right) =\\
& = & -3 D\left( (DN)^2 \right).
\end{eqnarray*}
All the anticommutators are proportional to the unit matrix (they are squares
of traceless $2\times 2$ matrices), so
\begin{eqnarray*}
\Phi _3 & = & -{1\over 32} \{DN,(DN)^2\} -{1\over 32} \{\1 , D^3N\} 
-{3\over 64} \left\{ N, D\left( (DN)^2 \right) \right\} 
- {1\over 64} (DN)^3 = \\
& = & -{5\over 64} (DN)^3 - {1\over 16} D^3N - 
{3\over 64} N D\left( (DN)^2 \right).
\end{eqnarray*}

To obtain the gradient expansion of the free energy up to the fourth
order, we need to calculate ${\cal L}_2$ and ${\cal L}_4$ from (\ref{Lrel}):
\begin{eqnarray*}
{\cal L}_2 & = & (DN) \Phi _1 = -{1\over 4} (DN)^2 \\
{\cal L}_4 & = & (DN) \Phi _3 = -{5\over 64} (DN)^4 - {1\over 16} (DN)\cdot (D^3N)-
{3\over 64} (DN)\cdot N\cdot D\left( (DN)^2 \right),
\end{eqnarray*}
so
\begin{eqnarray*}
L_2 & = & -{\zeta \over 8} \tr \left( (DN)^2 \right) \\
L_4 & = & -{5\over 128} \zeta \tr \left( (DN)^4 \right) -
{1\over 32}\zeta \tr \left( (DN)\cdot (D^3N) \right) -
{3\over 64} \zeta \tr \left( (DN)\cdot N\cdot D\left( (DN)^2 \right) \right).
\end{eqnarray*}
We can rewrite $L_4$ in a more symmetric way by writing
$$
\zeta \tr \left( (DN)\cdot (D^3N) \right) = 
\zeta D\left( \tr \left( (DN)\cdot (D^2N) \right) \right) - 
\zeta \tr \left( (D^2N)^2 \right). 
$$
The first term is a total derivative
($\zeta D = \partial$ by definition (\ref{D})), 
and is, therefore thrown out. The
last term in $L_4$ is equal to zero too, since 
$$
\tr \left( (DN)\cdot N \right) = {1\over 2} \tr \left( \{DN,N\} \right) = 0,
$$
and $D \left( (DN)^2 \right)$ is proportional to the unity matrix.
Thus, up to the fourth order in derivatives, including even-order
terms only,
\begin{equation}
\label{lagr024}
L \equiv L_0+ L_2 + L_4 = \zeta \tr \left\{ \1
-{1\over 8} (DN)^2 - {5\over 128} (DN)^4 + {1\over 32} (D^2 N)^2
\right\}.
\end{equation}

\section{The Free Energy}

The equation (\ref{lagr024}) at the end of the previous section is
the principal result of this paper.  It contains, in an extremely
compact form, all the information needed to obtain the free energy.
Using the results of section 2 this may be written
\begin{equation}
\label{EQ:finalF}
{\cal F} = \int d^3x {|{\bf H - H }_a |^2 \over 8\pi } + 
\int d^3x {|\Delta |^2 \over V } + 2\pi N_0 \int 
{d\Omega _{\bf n}\over 4\pi }\int d^2x_{\perp } \f1d ,
\end{equation}
where $\f1d $ is given in terms of $L$ as 
\begin{equation}
\f1d = -{T\over 2 } \sum\limits_{\omega _m} \int dx_{\parallel} L .
\end{equation}
Since $d^2x_{\perp } dx_{\parallel} = d^3x , $ we can write
\begin{equation}
\label{ffull}
{\cal F} = \int d^3r \left\{ {|{\bf H - H }_a |^2 \over 8\pi } +
{|\Delta |^2 \over V } - \pi T N_0 \sum\limits_{\omega _m}
{d\Omega _{\bf n}\over 4\pi } L \right\}.
\end{equation}
Notice that terms with an odd number of derivatives, and hence an odd number
of vectors $\bf n$, drop out when we average over them. This explains
why we calculated only even terms in the expansion of $L$.

We have included a magnetic field in (\ref{EQ:finalF}).  In our
earlier calculations, we did not mention the coupling of the
electrons to the magnetic potential $\bf A$. However the  component
$A_{\parallel}$ can always be gauged away along the line ${\bf
x}_{\perp}= {\rm const.}$ by absorbing it into $\theta$.  There is
therefore no loss of generality in our formulae. To insert $\bf A$ we
merely replace our derivative of the complex order parameter by a
covariant derivative, and our derivative of the (real) magnitude of
the order parameter by a plain gradient.

We would like to compare our calculation with that of reference
\cite{tewordt} as this seems to be the only place where the fourth-order
terms have been written down explicitly.

The expression given for the free energy in \cite{tewordt}
is, in the notation of that paper, 
\begin{eqnarray}
\label{fTew}
{\cal F} & \stackrel{?}{=} & \int d^3r 
\biggl\{ {|{\bf H - H }_a |^2 \over 8\pi } +
N_0 \beta ^{-2} \Bigl[ \beta ^2 w + {1\over 2} (v_F \beta )^2
\bigl\{ g |{\bf O } \chi |^2 + {1\over 6} g' (\nabla |\chi |^2 )^2 
\bigr\} + \nonumber \\
& + & {1\over 12} (v_F \beta )^4 \bigl\{ g' | {\bf O}^2 \chi | ^2 +
g'' [{1\over 2} |{\bf O } \chi |^4 - |{\bf O } \chi |^2 
(\nabla ^2 |\chi |^2 ) + {1\over 10} (\nabla ^2 |\chi |^2 )^2] + 
\nonumber \\
& + & g''' [{1\over 4} (\nabla ^2 |\chi |^2 )(\nabla |\chi |^2 )^2 -
{1\over 2} |{\bf O } \chi |^2 (\nabla |\chi |^2 )^2 ] \bigr\}
\Bigr] \biggr\}.
\end{eqnarray}

Here $\chi = \beta \Delta $ is the dimensionless order parameter, and
${\bf O} = \nabla - 2ie \bf A$ is the covariant derivative.  The
homogeneous part of the free energy comes from $w(|\chi|)$, while $g$
is a function of  $ |\chi | ^2 $ that we will identify later ($g$ is
not to be confused with the Eilenberger function). The primes on $g$
denote the derivative with respect
 to $|\chi | ^2$, {\it i.e.\/}
$$
g' \equiv {dg \over d |\chi | ^2 }.
$$
The directional averaging $\int {d\Omega _{\bf n}\over 4\pi }$ is not
stated explicitly  in this  formula, but is to be understood. In
other words the product of two expresions ${\bf  A}{\bf B}$ where
${\bf  A}$, ${\bf B}$ are either of the two vector operators $\bf O$
or $\nabla$ is defined as
\be
{\bf  A}{\bf B} =\frac 13 \delta_{ij} A_i B_j,
\ee
and the product of  four factors as
\be
{\bf  A}{\bf B}{\bf  C}{\bf D}= 
\frac 1{15}(\delta_{ij}\delta_{kl}+\delta_{ik}\delta_{lj}
+\delta_{il}\delta_{jk})A_iB_jC_kD_l.
\ee 
 We have inserted a question mark over the
equals sign in (\ref{fTew}), because we believe that there is an error in the
fourth-order terms.

We will now  expand  our expression so as to write it in Tewordt's form.
During the calculation, we will keep the primes as derivatives, and
only at the end we will replace them by the covariant derivatives or
gradients according to the rule
\begin{eqnarray}
\label{rule}
\chi ' & \rightarrow & {\bf O} \chi \nonumber \\
{(|\chi |^2 )}' & \rightarrow &  \nabla |\chi |^2 .
\end{eqnarray}
Note that the
prime over the order parameter means $d/dx$, whereas the prime over
$g$ means $d/d|\chi |^2 $.

We introduce the dimensionless
quantities
\begin{equation}
\label{TewM}
{\cal M} \equiv \beta M = \pmatrix{ (2m+1) \pi & -i\chi \cr
i \chi ^* & - (2m+1) \pi }
\end{equation}
and
\begin{equation}
\label{xi}
\xi \equiv \beta \zeta = 
\sqrt{ (2m + 1)^2 \pi ^2 + |\chi |^2 },
\end{equation}
so
\begin{equation}
\label{TewN}
N={{\cal M}\over \xi }.
\end{equation}
Since Tewordt writes out the Fermi velocity $v_F$ explicitly (in our
calculation, we set $v_F =1$), the dimensionless derivative $D$ equals
\begin{equation}
\label{TewD}
D={v_F \over \zeta} \partial = {v_F \beta \over \xi } \partial .
\end{equation}

From (\ref{ffull}), the second-order term in $\cal F$ is 
\begin{eqnarray}
\label{f2}
{\cal F}_2 & = & {\pi T N_0 \over 8} \int d^3r 
\sum\limits_{\omega _m} \zeta \tr (DN)^2 = \nonumber \\
& = & N_0 \beta ^{-2} {1\over 2} (v_F \beta ) ^2 \int d^3r 
\sum\limits_{\omega _m} {\pi \over 4 } \xi
{1\over (v_F \beta ) ^2} \tr (DN) ^2 ,
\end{eqnarray}
where we pulled out the prefactors that appear in the Tewordt
formula. Using (\ref{TewN}) and (\ref{TewD}), we get
$$
{1\over v_F \beta } DN = 
{1\over \xi } \left( {{\cal M } \over \xi } \right) ' =
{{\cal M} ' \over \xi ^2} - {{\cal M} \xi ' \over \xi ^3} .
$$
To obtain the result in Tewordt's format, we need to trade the
derivatives of $\xi $  for the derivatives of $|\chi |^2 $.
From (\ref{xi}), we see that
\begin{equation}
\label{xiprime}
\xi ' = {{(|\chi |^2)}' \over 2\xi },
\end{equation}
so
\begin{equation}
\label{DN}
{1\over v_F \beta } DN = {{\cal M} \over \xi ^2} -
{{\cal M} {(|\chi |^2)}' \over 2 \xi ^4} .
\end{equation}
Hence,
$$
{1\over (v_F \beta )^2}(DN)^2 =
{{{\cal M}'}^2\over \xi ^4} - 
\{{\cal M}, {\cal M}'\} {{(|\chi |^2)}' \over 2 \xi ^6} +
{{\cal M}^2 {{(|\chi |^2)}'}^2 \over 4\xi ^8} .
$$
From (\ref{TewM}) we see
\begin{eqnarray}
\label{M2}
{\cal M}^2 & = & \xi ^2 \1 \\
\{{\cal M}, {\cal M}' \} & = & {({\cal M })^2}' =
{(|\chi |^2)}' \1 \\
\label{Mprime2}
{{\cal M}'}^2 & = & \chi ' {\chi ^*}' \1 ,
\end{eqnarray}
so
\begin{equation}
\label{DN2}
{1\over (v_F \beta )^2} (DN)^2  =  \left(
{\chi ' {\chi ^* }' \over \xi ^4} -
{{{(|\chi | ^2)} ' }^2 \over 4 \xi ^6} \right) \1 .
\end{equation}
We now use (\ref{rule}) to replace the derivatives, and
obtain
\begin{equation}
{\cal F}_2 = N_0 \beta ^{-2} {1\over 2} (v_F \beta )^2 
\int d^3r \sum\limits_{\omega _m} {\pi \over 2} \left(
{|{\bf O }\chi |^2 \over \xi ^3 }-
{{(\nabla |\chi |^2 )}^2 \over 4\xi ^5 } \right) .
\end{equation}
Comparing the first term to (\ref{fTew}), we see immediately
\begin{equation}
\label{gTew}
g={\pi \over 2} \sum\limits_{\omega _m} {1\over \xi ^3} .
\end{equation}
We will also need higher derivatives of $g$ with respect to
$|\chi |^2$. Using (\ref{xi}), we get
\begin{eqnarray}
\label{gderivatives}
g' & = & -{3\over 4} \pi \sum\limits_{\omega _m}
{1\over \xi ^5} \nonumber \\
g'' & = & {15\over 8} \pi \sum\limits_{\omega _m}
{1\over \xi ^7} \nonumber \\
g''' & = & -{105 \over 16} \pi \sum\limits_{\omega _m}
{1\over \xi ^9} .
\end{eqnarray}
Hence,
\begin{equation}
{\cal F}_2 = N_0 \beta ^{-2} {1\over 2} (v_F \beta )^2 
\int d^3r \left( g |{\bf O }\chi |^2 + 
{g'\over 6} {(\nabla |\chi |^2 )}^2 \right)
\end{equation}
in agreement with (\ref{fTew}).

The fourth-order term is equal to
\begin{eqnarray}
\label{f4}
{\cal F}_4 & = & -\pi T N_0 \int d^3r \sum\limits_{\omega _m}
\zeta \tr \left( -{5\over 128} (DN)^4 + 
{1\over 32} (D^2N)^2 \right) = \nonumber \\
& = & N_0 \beta ^{-2} {(v_F \beta )^4 \over 12} \int d^3r
\sum\limits_{\omega _m} {\xi \pi \over (v_F \beta )^4} \tr 
\left( {15 \over 32} (DN)^4 - {3\over 8} (D^2N)^2 \right) .
\end{eqnarray}
We can calculate the first term immediately from (\ref{DN2})
\begin{equation}
\label{DN4}
{1 \over (v_F \beta )^4} (DN)^4 = \left(
{(\chi ' {\chi ^*}')^2 \over \xi ^8} -
{\chi ' {\chi ^*}' {{(|\chi |^2)}'}^2 \over 2 \xi ^{10}} +
{ {{(|\chi |^2)}'}^4 \over 16 \xi ^{12}} \right) \1 .
\end{equation}
To evaluate the second term in (\ref{f4}), we need to go
back to (\ref{DN}):
\begin{eqnarray*}
{1 \over (v_F \beta )^4} D^2N & = & 
{1\over \xi} {\left( {{\cal M}' \over \xi ^2} -
{{\cal M} {(|\chi |^2 )}' \over 2 \xi ^4 }\right)}' = \\
& = & {{\cal M}''\over \xi ^3} 
-{3\over 2} {{\cal M}' {(|\chi |^2 )}' \over \xi ^5 }+
{\cal M} \left( { {{(|\chi |^2)}'}^2\over \xi ^7 }-
{{(|\chi |^2)}'' \over 2 \xi ^5 } \right)
\end{eqnarray*}
(we used again (\ref{xiprime}) to obtain the second line), so
\begin{eqnarray*}
{1 \over (v_F \beta )^4} (D^2N)^2 &=&
{{{\cal M}''}^2 \over \xi ^6 }+ 
{9\over 4} {{{\cal M}'}^2 {{(|\chi |^2)}'}^2 \over \xi ^{10} }+
{\cal M} ^2 {\left( { {{(|\chi |^2)}'}^2\over \xi ^7} -
{{(|\chi |^2)}'' \over 2 \xi ^5 } \right) }^2 - \\
& - & {3\over 2} \{{\cal M}'', {\cal M}' \}{{(|\chi |^2)}' \over \xi ^8}
-{3\over 2}  \{{\cal M}', {\cal M} \}{(|\chi |^2)}'
\left( { {{(|\chi |^2)}'}^2\over \xi ^{12}} -
{{(|\chi |^2)}'' \over 2 \xi ^{10} } \right) + \\
& + & \{{\cal M}'', {\cal M}\} \left( { {{(|\chi |^2)}'}^2\over \xi ^{10}} -
{{(|\chi |^2)}'' \over 2 \xi ^8 } \right)
\end{eqnarray*}
From (\ref{TewM}), (\ref{M2})-(\ref{Mprime2}), we see that
\begin{eqnarray*}
{{\cal M}''}^2  & = & \chi '' {\chi ^*}'' \1 \\
\{{\cal M}'', {\cal M}' \} & = & {({{\cal M}' }^2)}' \1 \\
\{{\cal M}'', {\cal M}\} & = & {\{{\cal M}', {\cal M}\}}' -
2{ {\cal M}'}^2 = ( {(|\chi |^2)}'' - 2 \chi ' {\chi ^* }' )\1 ,
\end{eqnarray*}
so
\begin{eqnarray*}
{1 \over (v_F \beta )^4} (D^2N)^2 &=& \biggl(
{\chi '' {\chi ^*}'' \over \xi ^6} +
{1\over 4}{\chi '{\chi ^*}' {{(|\chi |^2)}'}^2\over \xi ^10} -
{1\over 2} {{{(|\chi |^2)}'}^4 \over \xi ^{12}} -
{{(|\chi |^2)}'' \over 4 \xi ^8 } - \\
& - & {3\over 2} {\chi ' {\chi ^*}}' {{(|\chi |^2)}' \over \xi ^8}+
{3\over 4} {{{(|\chi |^2)}'}^2 {(|\chi |^2)}'' \over \xi ^{10}} +
{\chi ' {\chi ^* }' {(|\chi |^2)}'' \over \xi ^8} \biggr)
\end{eqnarray*}
Putting the two terms together gives
\begin{eqnarray*}
{\cal F}_4 & = & N_0 \beta ^{-2} {(v_F \beta )^4 \over 12} \int d^3r
\sum\limits_{\omega _m} \pi \biggl( 
{15 \over 16} {{\chi ' {\chi ^*}'}^2 \over \xi ^7 }-
{21\over 32} {\chi ' {\chi ^*}' {{(|\chi |^2)}'}^2 \over \xi ^9} +
\underline{{111\over 16^2}{ {{(|\chi |^2)}'}^4 \over \xi ^{11}}} - \\
& - & {3\over 4 }{\chi '' {\chi ^*}'' \over \xi ^5} +
{3\over 16} {{{(|\chi |^2)}''}^2 \over \xi ^7 }+
\underline{ {9\over 8}{(\chi ' {\chi ^*}')}' 
{{(|\chi |^2)}' \over \xi ^7}} -
{9\over 16} {{{(|\chi |^2)}'}^2 {(|\chi |^2)}'' \over \xi ^9}-
{3\over 4} {\chi ' {\chi ^* }' {(|\chi |^2)}'' \over \xi ^7} \biggr).
\end{eqnarray*}
The two underlined terms do not appear in Tewordt's formula, so we
need to integrate them by parts (again neglecting the boundary
terms):
\begin{eqnarray*}
{111\over 16^2}{ {{(|\chi |^2)}'}^4 \over \xi ^{11}} & = & 
{111\over 16 \times 8} {{(|\chi |^2)}'}^3 {1\over \xi ^{10}}
{{(|\chi |^2)}' \over 2\xi } = \\
& = & -{1\over 9}{111\over 16 \times 8} {{(|\chi |^2)}'}^3
{\left( {1\over \xi ^9} \right) }' = \\
& = & -{1\over 3} {37 \over 16 \times 8 } {\left(
{{(|\chi |^2)}'}^3{1\over \xi ^9} \right) }' +
{37\over 16\times 8} {{{(|\chi |^2)}'}^2 {(|\chi |^2)}'' \over \xi ^9}
\end{eqnarray*}
and
\begin{eqnarray*}
{9\over 8} {(\chi ' {\chi ^*}')}' {{(|\chi |^2)}' \over \xi ^7} & = &
{9\over 8} {\left( \chi ' {\chi ^*}' {{(|\chi |^2)}' \over \xi ^7}
\right) }' -
{9\over 8} \chi ' {\chi ^*}' {{(|\chi |^2)}'' \over \xi ^7} +
{63\over 16}{\chi ' {\chi ^*}' {{(|\chi |^2)}'}^2 \over \xi ^9}
\end{eqnarray*}
Therefore,
\begin{eqnarray*}
{\cal F}_4 & = & N_0 \beta ^{-2} {(v_F \beta )^4 \over 12} \int d^3r
\sum\limits_{\omega _m} \pi \biggl( {15 \over 16} 
{{\chi ' {\chi ^*}'}^2 \over \xi ^7 }
+ {105\over 32} {\chi ' {\chi ^*}' {{(|\chi |^2)}'}^2 \over \xi ^9}-
{3\over 4} {\chi '' {\chi ^*}'' \over \xi ^5} + \\
& + & {3\over 16} {{{(|\chi |^2)}''}^2 \over \xi ^7 }-
{35\over 16\times 8} {{{(|\chi |^2)}'}^2 {(|\chi |^2)}'' \over \xi ^9}
-{15\over 8} {\chi ' {\chi ^* }' {(|\chi |^2)}'' \over \xi ^7} \biggr).
\end{eqnarray*}
Using (\ref{gderivatives}) and introducing the covariant derivatives and
gradients, we finally get
\begin{eqnarray}
\label{EQ:tewordt4}
{\cal F}_4 & = & N_0 \beta ^{-2} {(v_F \beta )^4 \over 12} \int d^3r
\biggl( g' | {\bf O}^2 \chi | ^2 +
g'' [{1\over 2} |{\bf O } \chi |^4 - |{\bf O } \chi |^2
(\nabla ^2 |\chi |^2 ) + {1\over 10} (\nabla ^2 |\chi |^2 )^2]
 + \nonumber\\
& + & g''' [
-{1\over 2} |{\bf O } \chi |^2 (\nabla |\chi |^2 )^2  
+{1\over 24} (\nabla ^2 |\chi |^2 )(\nabla |\chi |^2 )^2 ] \biggr).
\end{eqnarray}
We see that our formula agrees with Tewordt's result except for the
prefactor in the last term (1/24 instead of 1/4). This is presumably
a typographical error.
We believe our
result is correct, because we had previously obtained it by two
other methods.

\section{Discussion}

While principal result of the paper is the fourth-order term in the
free energy, as expressed in equations (\ref{lagr024}) and
(\ref{EQ:tewordt4}), the methods used to obtain this term are of
interest in their own right. The useful identity linking the dressing
function $\phi$ with the determinant  is one that we have not seen
before, and the purely algebraic (requiring no integration)
generation of the terms is conceptually simpler than other methods of
obtaining such series such as those we used in \cite{kosztin}. In
particular the present algorithm produces very compact expressions,
and is applicable to the computation of determinants of any $2\times
2$ first-order matrix operator.

\vspace*{-2pt}
\section*{Acknowledgments}

This   work  was supported  by the National Science Foundation under
grant number DMR94-24511 and by the University of Illinois  Science
and Technology Center for Superconductivity   under grant number
NSF-DMR-9120000.  We would like to thank Ioan Kosztin for valuable
converations. MS would like to thank Gerald Dunne for sending him
some of his unpublished notes on the Gelfand-Dikii equation for
periodic boundary conditions, and David Waxman for e-mail
discussions.


\vspace*{-9pt}

\vfil
\eject
\end{document}